\documentclass[aps,apl,amsmath,amssymb,reprint,superscriptaddress]{revtex4-1}
\usepackage{graphicx}
\usepackage{amsmath}
\usepackage{dcolumn}
\usepackage{bm}
\usepackage{bbold}
\usepackage{color}
\usepackage{amsfonts}
\usepackage{amssymb}
\usepackage{mathrsfs}
\usepackage{tabularx}
\usepackage{braket}
\usepackage{mathtools}
\usepackage{soul}			

\usepackage{caption}
\usepackage{subcaption}

\begin{document} 
\title{Determination of Low Loss in Isotopically Pure Single Crystal $^{28}$Si at Low Temperatures and Single Microwave Photon Energy}

\author{Nikita Kostylev}
\affiliation{ARC Centre of Excellence for Engineered Quantum Systems, School of Physics, University of Western Australia, 35 Stirling Highway, Crawley WA 6009, Australia}

\author{Maxim Goryachev}
\affiliation{ARC Centre of Excellence for Engineered Quantum Systems, School of Physics, University of Western Australia, 35 Stirling Highway, Crawley WA 6009, Australia}

\author{Andrey D. Bulanov}
\affiliation{G.G. Devyatykh Institute of Chemistry of High-Purity Substances of the Russian Academy of Sciences, 49 Tropinin Str., Nizhny Novgorod 603950, Russia}
\affiliation{N.I. Lobachevski State University, 23 Gagarin prosp., Nizhny Novgorod 603950, Russia}

\author{Vladimir A. Gavva}
\affiliation{G.G. Devyatykh Institute of Chemistry of High-Purity Substances of the Russian Academy of Sciences, 49 Tropinin Str., Nizhny Novgorod 603950, Russia}

\author{Michael E. Tobar}
\email{michael.tobar@uwa.edu.au}
\affiliation{ARC Centre of Excellence for Engineered Quantum Systems, School of Physics, University of Western Australia, 35 Stirling Highway, Crawley WA 6009, Australia}

\date{\today}


\begin{abstract}

The low dielectric losses of an isotropically pure single crystal $^{28}$Si sample were determined at a temperature of 20 mK and at  powers equivalent to that of a single photon. Whispering Gallery Mode (WGM) analysis revealed large Quality Factors of order $2\times10^6$ (dielectric loss $\sim 5\times10^{-7}$) at high powers, degrading to $7\times10^5$ (dielectric loss $\sim 1.4\times10^{-6}$) at single photon energy. A very low-loss narrow line width paramagnetic spin flip transition was detected with extreme sensitivity in $^{28}$Si, with very small concentration below $10^{11}$~cm$^{-3}$ (less than 10 parts per trillion) and g-factor of $1.995\pm0.008$. Such determination was only possible due to the low dielectric photonic losses combined with the long lifetime of the spin transition (low magnetic loss), which enhances the magnetic AC susceptibility. Such low photonic loss at single photon energy combined with the narrow line width of the spin ensemble, indicate that single crystal $^{28}$Si could be an important crystal for future cavity QED experiments.
\end{abstract}

\maketitle

Low loss crystals find a broad range of applications in many different areas of physics and engineering. One major purpose is the realization of high finesse and $Q$-factor photonic and phononic systems allowing resonances with extremely narrow frequency linewidths. For example, such utilisation is of fundamental importance for the best stability frequency sources\cite{Locke:2008aa, Kessler:2012}, fundamental physics tests \cite{Herrmann:2009, Nagel:2015}, gravitational wave detection \cite{Khalaidovski:2013, PhysRevD.90.102005} and opto-mechanical systems \cite{PhysRevLett.96.173901, PhysRevLett.108.033602}. Another example is quantum electrodynamics involving dielectric crystals where quantum signal coherence, one of the most important parameters, is related to various loss mechanisms and thus on crystal properties\cite{PhysRevB.90.100404,PhysRevA.82.033810,PhysRevLett.107.060502,Abe:2011aa,Baibekov:2011aa,Goryachev:2015ax,Farr:2015aa}, for these applications the properties at the energy level of a single photon is of paramount importance. In all these fields, special attention has been devoted to Silicon due to its excellent optical and mechanical properties and abundance of associated technologies for growth, treatment, implantation, etc. Furthermore, there is a constant need for new approaches and new materials due to emerging applications. In particular, a recent development in ion implementation with applications for quantum signal processing\cite{Lyon,Morton:2011aa,Pla:2012aa,Itoh:2014aa} induced significant interest in isotopically purified silicon crystals due to its unique properties. 

Among three stable isotopes of Silicon with 28, 29 and 30 nucleons (approximately 0.92\%, 0.04\% and 0.03\% abundance in nature), $^{28}$Si and $^{30}$Si have zero nuclear spin. This particular feature makes associated crystals act as a virtual vacuum where signals or implemented ions are not coupled to a bath of lattice magnetic Two Level Systems (TLS) leaving the medium inert \cite{Wolfowicz:2012}. Thus, absence of nuclear spin in the lattice surpasses one of the most important channels of decoherence leaving only phonon dissipation that can be greatly reduced by cooling down the crystals. In this work, we investigate low temperature microwave properties of $^{28}$Si single crystal down to the single photon level and undertake Electron Spin Resonance Spectroscopy.

\begin{figure}[t!]
			\includegraphics[width=0.45\textwidth]{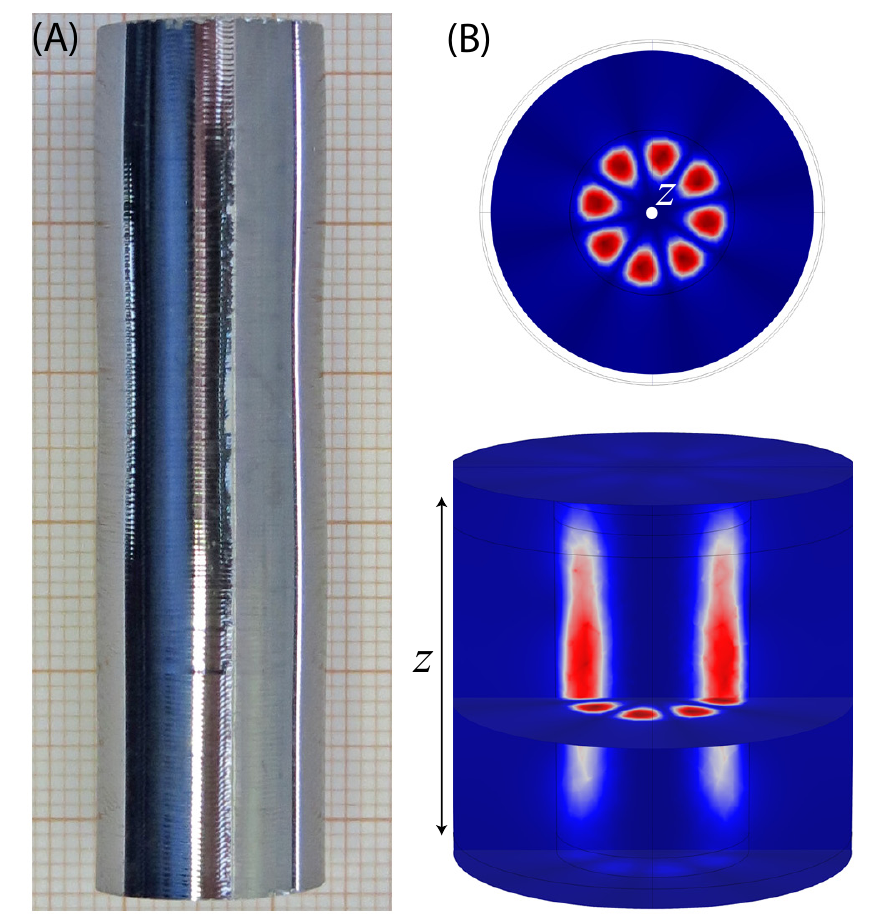}
	\caption{(A) Photo of the $^{28}$Si crystal sample. (B) Simulated field density for a WGM with of azimuthal mode number $m = 4$ and $\omega_0/2\pi$ = 7.01 GHz. }
	\label{spec1}
\end{figure}

Two samples of Silicon were investigated using the Whispering Gallery Mode (WGM) method. The first sample is a single crystal of isotopically purified $^{28}$Si cylinder of $H = 55$~mm and $D=15$~mm, see Fig.~\ref{spec1}, (A). This sample was grown at G.G.~Devyatykh Institute of Chemistry of High-Purity Substances using the float zone melting method in Argon atmosphere\cite{Itoh:2003aa}. The specified concentration of $^{28}$Si is 99.993\%, with concentration of Oxygen and Carbon estimated to be less than $1.1\times 10^{16}$ cm$^{-3}$ (on the order of a ppm) and Boron and Phosphorous to be less than $1.1\times 10^{13}$ cm$^{-3}$ (on the order of a ppb). 

The second sample is float-zone high-resistivity single crystal silicon with a natural abundance of its isotopes. It is cut in a form of a cylinder with height $H = 6$~mm and diameter $D=15$~mm and crystallographic direction (100). The crystal has been mechanically polished to reduce surface resistance but the finish is below optical quality. Microwave properties of this sample have already been characterized for a temperature range of 10-400 K \cite{Krupka:2006MP}.  It has been found that below 35 K the dielectric losses in the sample were decreasing, with Quality Factors reaching $10^{4}$, suggesting existence of hopping or surface charge carrier conductivities. From our measurements, our work has shown good consistency with the losses at 10 K, with the best confined WGMs achieving a similar $Q$-factor. Furthermore the permittivity of the sample was determined to be 11.45 at 10 K \cite{Krupka:2006MP}, from our simulations, both our samples (natural and isotopically pure) were found to be in good agreement with this value at 20 mK. 

 Both crystals were thoroughly cleaned in an acid bath and using purified methanol before being enclosed in their respective Oxygen Free Copper Cavities. The natural Silicon sample was fixed through its middle point via a Copper and dielectric posts. This type of clamping removes dissipation via the conducting surfaces due to the fact that the energy density of a WGM are concentrated on the outer edges of the dielectric sample but not inside. The WGMs were excited and detected using two straight antenna probes located on the top of the cavity.

\begin{figure}[t!]
			\includegraphics[width=0.45\textwidth]{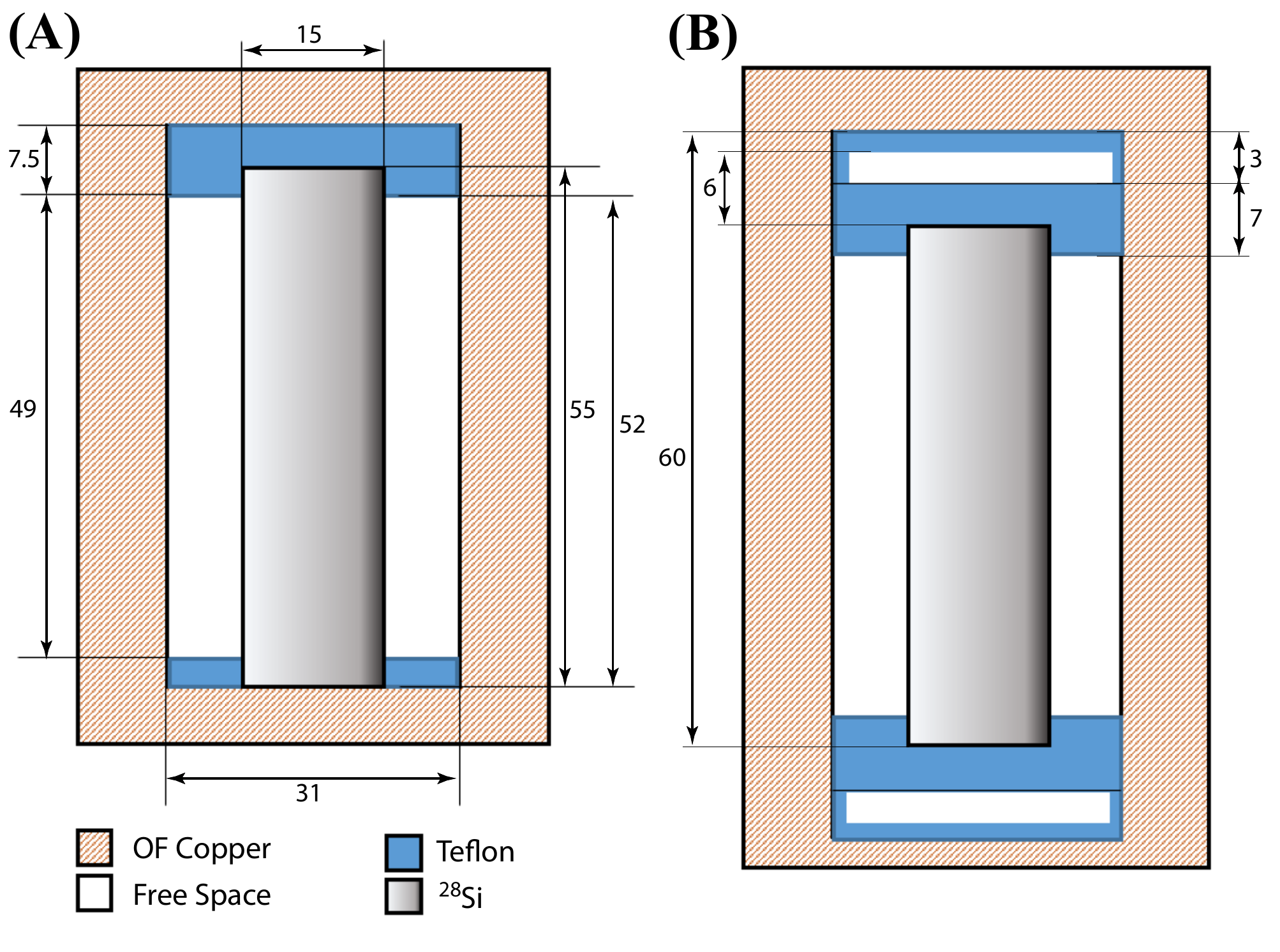}
	\caption{Schematic diagram of a simple (A) and Bragg reflector-type (B) WGM cavities used with the $^{28}$Si sample. All dimensions are given in millimeters.}
	\label{schem1}
\end{figure}

The other sample was mounted on a teflon dielectric. Because of the absence of the middle hole, two cavity designs were implemented and are shown in Fig. \ref{schem1}. The first design (Fig.~\ref{schem1}, (A)) provided a large contact area between one of the sample surfaces and copper walls to insure a good thermal connection  at the expense of adding metallic loss, which degrade the $Q$-factor. However, the long aspect ratio due to the large length of the sample relative to its diameter does reduce the metallic losses somewhat for WGMs compared to a smaller length sample. The second design (see Fig.~\ref{schem1}, (B)) implements Bragg reflectors to decouple modes of quasi-Transverse Electric (TE) polarization  from the metallic end walls\cite{Floch:2007aa,Floch:2008aa,Tobar:2001aa,Tobar:2004aa,Krupka:2005aa}, simultaneously improving the filling factor and reducing the metallic losses at the expense of a poor thermal contact. For both these configurations, the WGM waves were excited using loop probes located in the middle of the cavity side walls.

Both samples were placed one at a time inside a bore of 7 T superconducting magnet and were cooled to 20 mK while in vacuum with a dilution refrigerator. The experimental setup was similar to one used for WGM spectroscopy of other materials\cite{PhysRevB.88.224426,quartzQ,Goryachev:2015ab}. In this setup, the incident signal was attenuated to the level of few photons by a chain of room temperature and cold attenuators, and the output signal was amplified by a series of cold high-electron-mobility transistors (HEMT) and room temperature amplifiers. 

Using the described setups, a number of microwave modes of both crystals were measured in the frequency range 7-25 GHz. The detected resonances are higher order modes of different order, with the highest $Q$-factor modes being the WGMs (confirmed with Finite Element Analysis). An example of a WGM is provided in Fig.~\ref{spec1}, (B). Note that the loop electrodes made for the  $^{28}$Si sample may optimally couple to either $H_\phi$ or $H_z$, or a combination of fields, allowing coupling to a range of hybrid modes and WGMs of quasi TM and TE polarization respectively. 
 
\begin{figure}[t!]
			\includegraphics[width=0.45\textwidth]{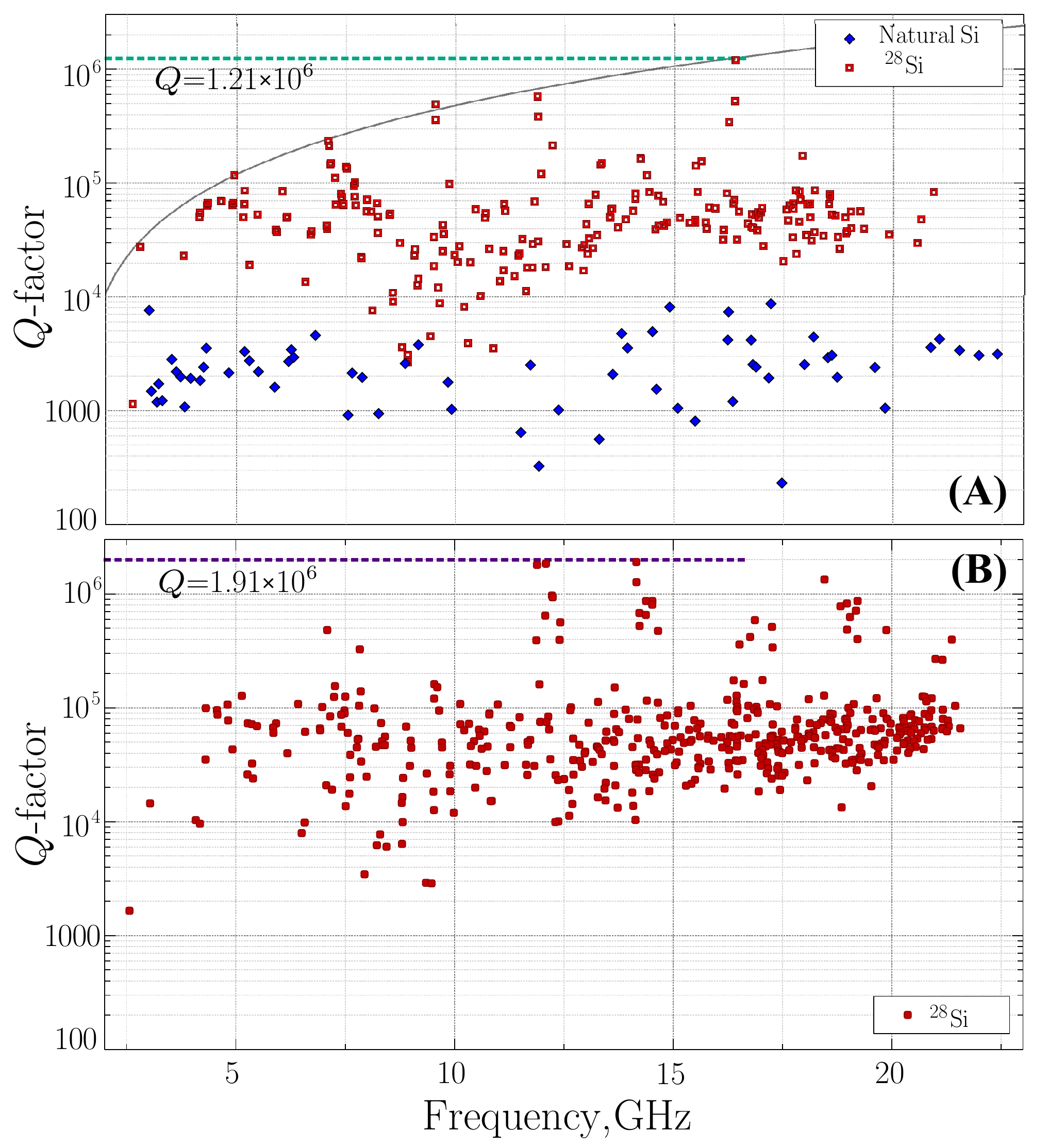}
	\caption{Quality Factor as a function of frequency for all measured electromagnetic modes in the natural silicon sample and the isotopically purified  $^{28}$Si crystal. (A) Blue diamonds show modes in the natural silicon cavity and red squares for the cavity in Fig. \ref{schem1} (A). The solid curve shows the upper limit associated with cavity metallic losses. (B) Quality Factors measured with a Bragg reflector-type cavity of Fig. \ref{schem1} (B). The measurements are for incident power of $P_{in}$=-60 dBm.}
	\label{Qfact}
\end{figure}

A large number of microwave modes in both X and K$_\text{u}$ frequency band allows us to establish microwave properties of the cavities as a function of frequency. In particular, the Quality Factor of the microwave modes shown in Fig.~\ref{Qfact} demonstrates photon losses in the material if the metallic losses can be made negligible. The measurements in Fig.~\ref{Qfact}, (A) are limited by the cavity wall losses due to the contact of the rod with the lid, with the WGMs exhibiting the highest $Q$-factors. Nevertheless, we come close to measuring the intrinsic material loss for a high-$Q$ mode at 16.4 GHz, at high enough frequency where the coupling to the cavity lid is significantly reduced. The results in Fig.~\ref{Qfact}, (B) provide a good indication of the intrinsic losses of the crystal due to the decoupling from the cavity walls via Bragg reflection ($Q$-factor $\sim 2 \times10^6$ for quasi TE WGMs between 10 to 15 GHz)). The Quality Factors in natural Silicon reach $10^4$ independent of frequency, which is in good agreement with earlier experiments by Krupka et. al.\cite{Krupka:2006MP}. 

\begin{figure}[t!]
			\includegraphics[width=0.5\textwidth]{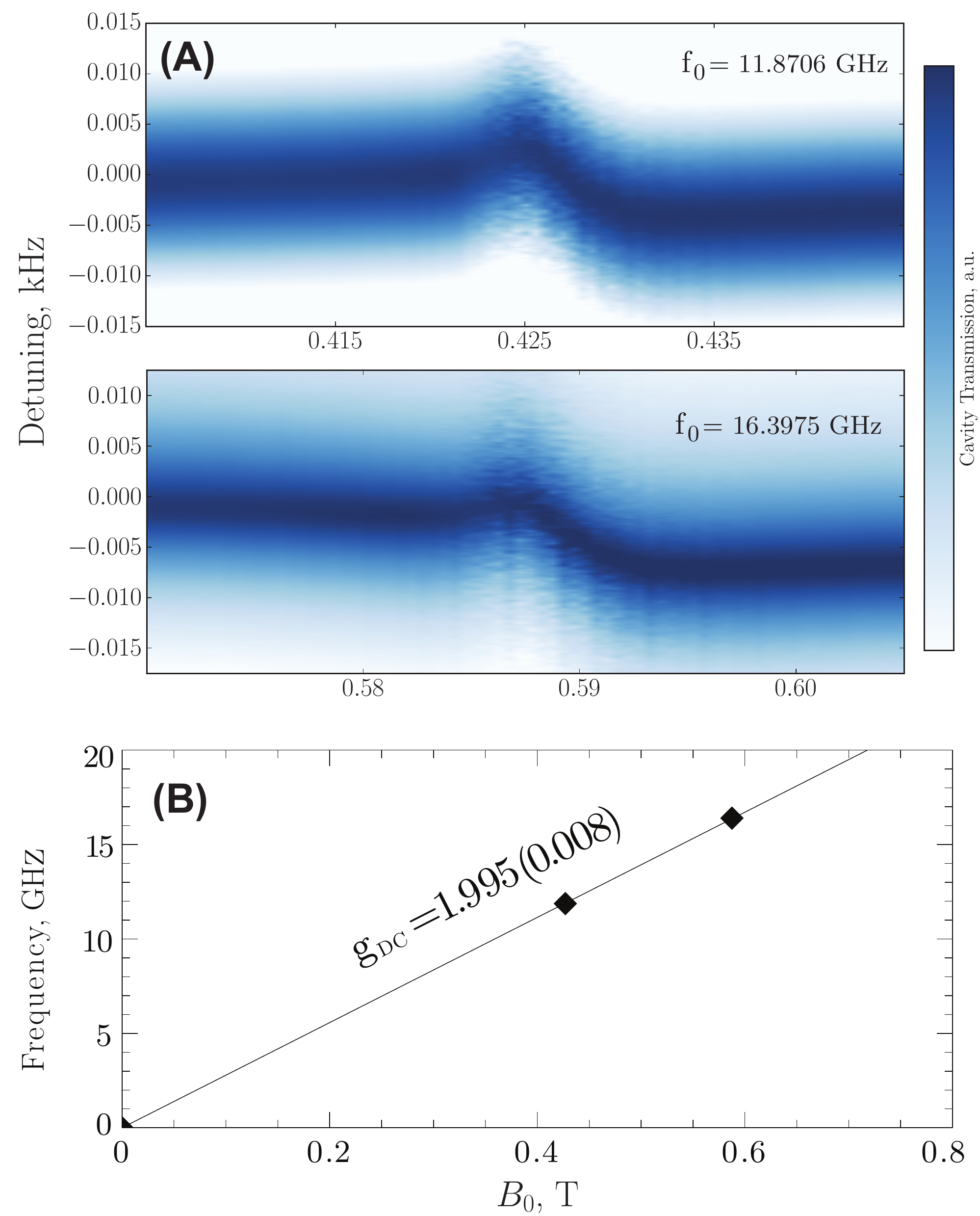}
	\caption{(A) Dependence of the crystal frequency response on the external magnetic field near the $11.87$ and $16.40$ GHz modes. (B) The detected points of interaction between the magnetic spin ensemble and the WG modes, assuming a spin flip transition that begins at the origin, an estimated Land\'{e} g-factor of $1.995\pm0.008$ is obtained from a linear fit with 2 sigma error.}
	\label{magfield}
\end{figure}

Information about crystal magnetic ion impurities may be obtained by monitoring the system microwave response while sweeping the external DC magnetic field $B_0$. Since none of the samples were intentionally doped, in this work, we only deal with residual impurities introduced during crystal growth. In both cases, $B_0$ was applied along the cylinder axis of symmetry. When the energy splitting of residual impurities swept by the external field reaches resonance frequency $f_0$ of a crystal electromagnetic mode, the shape of this mode is altered, as it is coupled to another harmonic oscillator with a close enough frequency. One interaction of such kind for the $^{28}$Si crystal is shown in Fig.~\ref{magfield}, (A). The plot does not display an Avoided Level Crossing between the ion ensemble and the WGM since the cavity photon linewidth $\delta$ is larger than the coupling, $g$, between the ion ensemble and the WGMs. However, only the highest $Q$-factor WGMs are sensitive enough to measure the transition, this is when the value of $g$ approaches the linewidth of the WGM.
In any case, each of such interactions can be mapped as a point onto a two dimensional $B(f)$ plane giving a larger scale picture of magnetic response of residual ions\cite{PhysRevB.88.224426}. Such a plot for the isotopically purified crystal is shown in Fig.~\ref{magfield}, (B), illustrating the existence of a spin-flip transition with an estimated Land\'{e} g-factor $g_\text{DC}$ of $1.995\pm0.008$. Note that no magnetic response have been detected for the natural Silicon due to significantly degraded photon Quality factors (observed experimentally) and potentially larger spin line widths arising because of nuclear spin of the crystal lattice.

The sensitivity of the method depends on the average ion concentration and spin transition lifetime $T_2$. The concentration can be determined from, $n=\frac{4\hbar}{\omega_0\mu_0\xi}\Big(\frac{g}{g_{DC}\beta_B}\Big)^2$, where $\omega_0$ is a cavity mode angular frequency, $\xi$ is the transverse magnetic filling factor, $\beta_B$ is the Bohr magnetron. Since $\delta>g$ (but of similar order for sensitive modes), $n$ cannot be accurately determined. However, the stated relation can give a upper bound for $n$ of the correct order of magnitude using the cavity half linewidth as the upper limit for the coupling $g$. Given that the filling factor of the mode, $\xi$ in  Fig.~\ref{magfield}, (B) is usually between 0.2 and 0.9 for the modes in the $^{28}$Si sample (as modelled using Finite Element analysis), an estimated concentration of  $10^{11}$ cm$^{-3}$ (on the order of ten parts per trillion) or less may be determined. Interestingly, the $Q$-factor of the WGM was found to remain constant as a function of the magnetic field throughout the photon-spin interaction. In most other crystals in which ESR resonances are broadened to more than a MHz the $Q$-factor will be significantly degraded. Thus the photon half linewidth also provides an upper bound of the spin transition linewidth, which is about 7 KHz. The actual spin linewidth is likely to be much lower than this, as very narrow spin linewidths have been measured previously, of order 10 Hz in $^{28}$Si \cite{Wolfowicz:2012}. It is this narrow spin linewidth, which enhances the AC susceptibility over the DC susceptibility by $\sim \omega_0 T_2$, which allows such a low concentration to be detected. Such narrow spin impurity line width is an important consequence of the absence of nuclear spins in the crystal lattice specific to the isotopically pure Silicon-28.

\begin{figure}[t!]
			\includegraphics[width=0.5\textwidth]{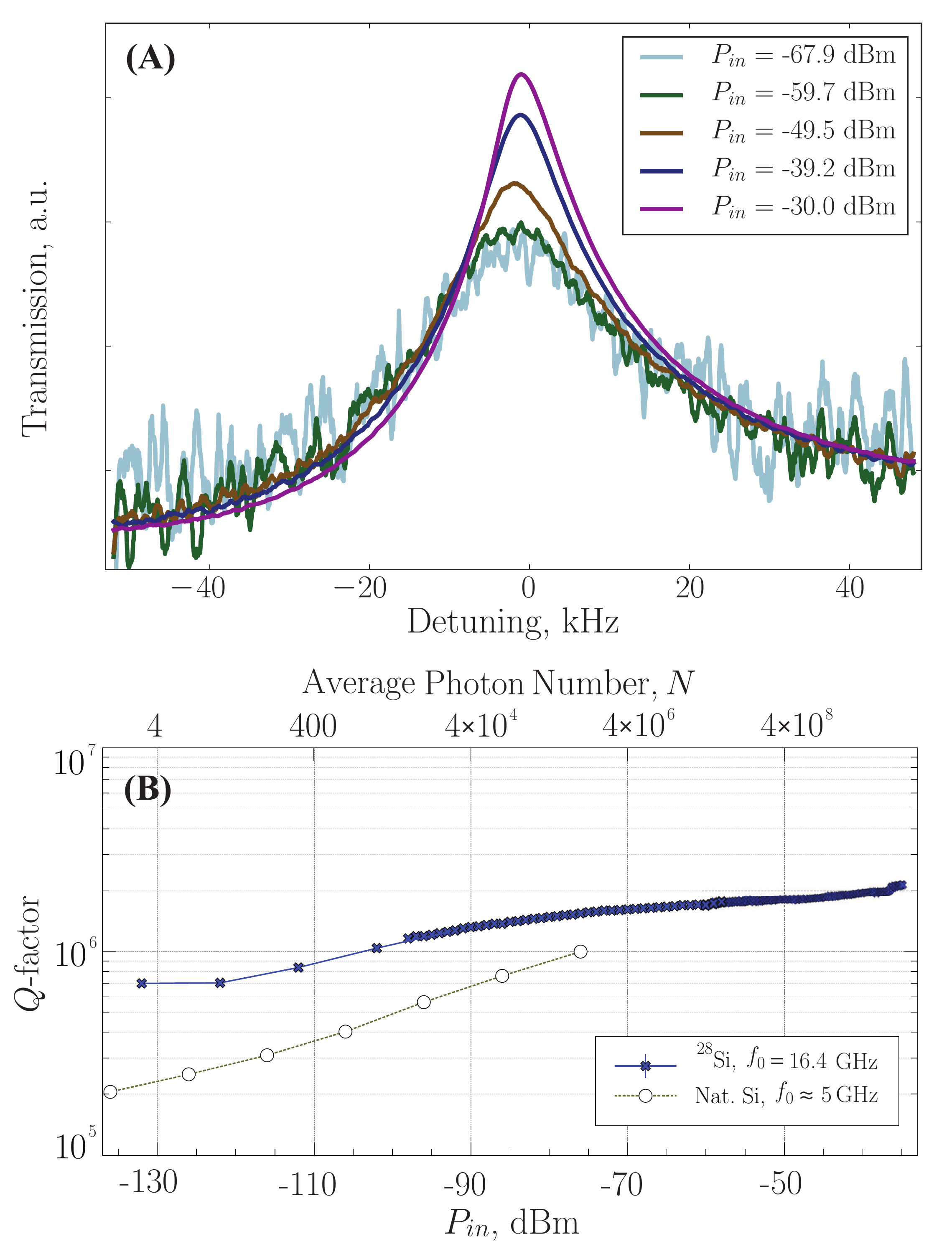}
	\caption{Power dependence of the $16.4$GHz mode. (A) Cavity transmission profile near resonance for selected incident powers, showing significant change in loss with little frequency shift. (B) Quality factor of the 16.4 GHz mode in isotopically-purified $^{28}$Si, compared to the $Q$-factor of a non-purified single crystal silicon capacitor lumped resonator at 5 GHz \cite{Weber:2011}. }
	\label{powerplot}
\end{figure}

As it has been specified above, the measurements were made down to very low levels of incident power (attenuated down to -132 dBm), required for a small number of excitations, with the average number of photons estimated to be on the order of $N=1$ to $2$ \cite{Hartnett:2011} . The results indicate that the isotopically-purified silicon has an advantage in terms of losses in the single-photon regime when compared to natural silicon. A $Q$-factor of $\approx7\times10^5$ (corresponding to a $Q\times f_0$ product $\sim 10^{16}$) was observed in our sample (Fig.~\ref{powerplot}, (C)), a factor of $3.5$ times higher than that reported in natural Si at $5$ GHz\cite{Weber:2011}. In terms of the $Q\times f_0$ product, the mode in the $^{28}$Si sample is a factor of 11.5 larger than that reported in natural Si. 

In general, the crystal microwave response shows nontrivial dependence on the incident power (see Fig.~\ref{powerplot}, (A, B) ), where transmission and mode Quality Factor grow with the increasing power. The dependence of $Q$ on incident power is different from the typical Kerr nonlinearity observed in many crystals, and is similar to that observed for rutile WGM resonators at 4K\cite{Nand:2014aa}, where the $Q$-factor increases with power with no corresponding frequency shift. For this case, such behavior was explained by WGM nonlinear losses originating from coupling to low frequency optical phonon modes. 

One can also attempt to explain the enhancement of $Q$ with increasing power by WGM coupling to a spin ensemble of TLS with considerable amount of dissipation. According to this model, a TLS ensemble becomes saturated at high incident power making the corresponding dissipation smaller compared to the overall circulating power\cite{Creedon:2011wk}. However, this mechanism induced by magnetic TLS has a narrow frequency range associated with the bandwidth of the ensemble. Moreover, we have shown in this work that the magnetic losses due to spins are extremely small. The current measurements are made at zero external magnetic field, where all existing magnetic impurities are detuned to zero frequency as shown in Fig.~\ref{magfield}, (B), ruling out the described mechanism in favour of a dielectric loss mechanism. 

In summary, we have investigated the low temperature properties of isotopically purified $^{28}$Si single crystals. The induced WGMs 
demonstrate higher Quality Factors than those observed in natural Silicon crystals. The magnetic field spectroscopy reveals small quantities of magnetic impurities with spin flip transitions with concentrations not exceeding $10^{11}$~cm$^{-3}$. The crystal also demonstrates some power dependence in the form of increasing Quality Factor.

We are grateful to Jerzy Krupka for supplying us with the natural silicon sample. This work was supported by Australian Research Council grant CE110001013.

\bibliography{biblio}

\end{document}